\RequirePackage{fix-cm}
\documentclass[twocolumn,epjc3]{svjour3}  
\smartqed  % flush right qed marks, e.g. at end of proof
\RequirePackage{graphicx}

\journalname{Eur. Phys. J.}
\usepackage[utf8]{inputenc}
\usepackage[british,UKenglish,USenglish,american]{babel}
\usepackage{amsmath}

\usepackage{graphicx}
\usepackage{subcaption}
\usepackage{slashed}

\usepackage{mathrsfs}
\usepackage{amssymb}
\usepackage{bbm}

\usepackage{nicefrac}

\usepackage{microtype}

\usepackage{physics}

\usepackage{xcolor}

\usepackage{dsfont}

\usepackage{widetext}

\newcommand\numberthis{\addtocounter{equation}{1}\tag{\theequation}}

\newcommand\operator{\hat{\mathcal{O}}}

\renewcommand{\bar}{\overline}

\usepackage{bm}\renewcommand{\vec}{\bm}

\newcommand{\e}{\mathrm{e}}

\newcommand{\CPV}{\mathrm{p.v.}}

\begin{document}

\title{
The strange critical endpoint and isentropic trajectories in an Extended PNJL Model with Eight Quark Interactions
}


\author{
Renan Câmara Pereira \thanksref{e1,addr1}
\and
João Moreira \thanksref{e2,addr1} 
\and
Pedro Costa \thanksref{e3,addr1}
}

\thankstext{e1}{e-mail: renan.pereira@student.uc.pt}
\thankstext{e2}{e-mail: jmoreira@uc.pt}
\thankstext{e3}{e-mail: pcosta@uc.pt}

\institute{
Centro de Física da Universidade de Coimbra (CFisUC), Department of Physics, 
Faculty of Sciences and Technology, University of Coimbra, P-3004 - 516  Coimbra, 
Portugal \label{addr1}
}

\date{Received: date / Accepted: date}

\maketitle

\begin{abstract}
In this work, we explore the possible existence of several critical endpoints in the phase diagram of strongly interacting matter using an extended PNJL model with 't Hooft determinant and eight quark interactions in the up, down and strange sectors. Besides, we also study the isentropic trajectories crossing both (light and strange) chiral phase transitions and around the critical endpoint in both the crossover and first-order transition regions.
\end{abstract}

%%%%%%%%%%%%%%%%%%%%%%%%%%%%%%%%%%%%%%%%%%%%%%%%%%%%%%%%%%%%%%%%%%%%%%%%%%%%%%%%
%%%%%%%%%%%%%%%%%%%%%%%%%%%%%%%%%%%%%%%%%%%%%%%%%%%%%%%%%%%%%%%%%%%%%%%%%%%%%%%%
\section{Introduction}
\label{intro}

The nature of the phase transition between hadron matter and the quark gluon plasma (QGP) and the search for the eventual QCD chiral critical endpoint (CEP) in the phase diagram of strongly interacting matter, have led to remarkable theoretical and experimental developments in recent years \cite{Halasz:1998qr,Brambilla:2014jmp}.
%The possible discovery of the CEP would be important in itself, but it would also lead to the answer on another central question that is whether the continuous chiral crossover that occurs at $\mu_B = 0$ \cite{Aoki:2006we} between hadronic matter, where the relevant degrees of freedom are hadronic, and the QGP, dominated by partonic degrees of freedom, becomes a ﬁrst-order phase transition above some critical point at a finite $\mu_B$.

The hypothetical discovery of the CEP would mean that the continuous temperature induced chiral crossover that occurs at $\mu_B = 0$ \cite{Aoki:2006we} between hadronic matter, where the relevant degrees of freedom are hadronic, and the QGP, dominated by partonic degrees of freedom, becomes a first-order phase transition above some critical point at a finite $\mu_B$.

%It is quite possible that detect the first-order phase-coexistence region would be easier than localize the CEP.
It is quite possible that the detection of the first-order phase-coexistence region is easier to achieve than that of the CEP.
% Indeed, when the expanding matter created in a heavy ion collision (HIC) pass through a putative first-order phase transition line,  the system probably will spend enough time in this region to develop measurable signals.}
Indeed, it is conjectured that when the expanding matter created in a heavy ion collision (HIC) passes through a putative first-order phase transition line, the system will probably spend enough time in this region to develop measurable signals \cite{Friman:2011zz}.

Fluctuation enhancement due to spinodal instabilities are a generic trait in systems undergoing a first-order phase transition, as such, despite the short lifetime and finite size of these systems, one can expect, as a possible telltale sign of this scenario, the appearance of enhanced fluctuations in the strangeness sector. These can result, for instance, in enhanced kaon-to-pion fluctuations (see \cite{Friman:2011zz} for a review).

%From the theoretical point of view, the phase diagram of strongly interacting matter, namely the possible location of the CEP, has been intensively studied by using lattice QCD (LQCD) calculations (despite the fermion sign problem, extrapolation methods were used to explore the region of small chemical potentials and look for the CEP \cite{Endrodi:2011gv}), and more recently by using Dyson-Schwinger equations \cite{Fischer:2018sdj,Isserstedt:2019pgx}. Also, effective models such as the Nambu–Jona-Lasinio (NJL) model and its extensions,  like the NJL model with eight-quark interactions and the Polyakov–Nambu–Jona-Lasinio (PNJL) model, have been used to study the CEP and critical properties around it \cite{Costa:2008yh,Costa:2008gr,Hiller:2008nu}. However, it remains unclear that such a CEP exists. The only theoretical solid evidence about the phase diagram comes from LQCD results at zero baryonic chemical potential that showed the existence of an analytic crossover from a low-temperature region characterized by chiral symmetry breaking to a high-temperature (partially) chirally restored region .The pseudocritical temperature for the chiral crossover is set at $T_c \approx 156$ MeV within a definition-dependent range of several MeV \cite{Borsanyi:2010bp,Bazavov:2011nk,Bellwied:2015rza,Bazavov:2018mes}.

%Acrescentei referencias e modifiquei texto

From the theoretical point of view, the phase diagram of strongly interacting matter, namely the possible location of the CEP, has been intensively studied by using lattice QCD (LQCD) calculations (despite the fermion sign problem, extrapolation methods were used to explore the region of small chemical potentials and look for the CEP \cite{Endrodi:2011gv}), and more recently by using Dyson-Schwinger equations \cite{Fischer:2018sdj,Isserstedt:2019pgx}. Effective models such as the Nambu–Jona-Lasinio (NJL) model \cite{Nambu:1961tp,Nambu:1961fr} and its extensions have also proven themselves to be invaluable tools in these explorations. Among these extensions, and of particular relevance for the present work, we highlight the inclusion of the $U_A(1)$ breaking 't Hooft flavor determinant \cite{'tHooft:1976fv,PhysRevD.18.2199.3} (introduced in the context of the NJL model in \cite{Kunihiro:1987bb,Bernard:1987sg,Reinhardt:1988xu}), the inclusion of eight-quark interactions \cite{Osipov:2005tq,Osipov:2006ns} and the Polyakov–Nambu–Jona-Lasinio (PNJL) model \cite{Fukushima:2003fw,Megias:2003ui,Megias:2004hj,Roessner:2006xn}. These have been used to study the phase diagram including the CEP and critical properties around it \cite{Hatsuda:1994pi,Zhuang:1994dw,Buballa:2003qv,Hansen:2006ee,Hiller:2008nu,Costa:2008yh,Costa:2008gr,Torres-Rincon:2017zbr}. 

It should be stressed however, that, at the present point, even the issue of the existence of such a CEP remains unclear. The only theoretical solid evidence about the phase diagram comes from LQCD results at zero baryonic chemical potential that showed the existence of an analytic crossover from a low-temperature region characterized by chiral symmetry breaking to a high-temperature (partially) chirally restored region. From LQCD, the pseudocritical temperature for the chiral crossover is found to be $T_c \approx 156~\mathrm{MeV}$ within a definition-dependent range of several MeV \cite{Borsanyi:2010bp,Bazavov:2011nk,Bellwied:2015rza,Bazavov:2018mes}.

Interestingly, some model calculations, namely in the NJL model with 't Hooft and eight-quark interactions supplemented with explicit chiral symmetry breaking interactions \cite{Osipov:2012kk,Osipov:2013fka} (those relevant in the sense of a $1/N_c$ expansion, with $N_c$ the number of colors) show that is possible for the strange sector to also have a first-order phase transition, meaning that a second CEP for this sector can also exist in the phase diagram \cite{Moreira:2014qna}. It should be noted that this secondary CEP does not appear in the phase diagram of the NJL with only 't Hooft and eight-quark interactions in \cite{Hiller:2008nu} but there the meson masses used to fit the model parameters \cite{Osipov:2006ns} are derived using a heat kernel expansion of Laplacian associated to the Dirac operator in the presence of background meson fields \cite{Osipov:2001nx,Osipov:2001bj,Osipov:2000rg}.
%\textbf{Falar mais: heat kernel e explicit chiral symmetry breaking interactions}
%\textbf{Reverificar: CEP2 HK3D}

As for the influence of external magnetic fields, besides affecting the location of the CEP \cite{Costa:2015bza}, it can lead to the appearance of the secondary CEP in the strange sector \cite{Ferreira:2017wtx}. Indeed, different versions of NJL and PNJL models have been used to study the location of the CEP. For example, the presence of repulsive vector interactions was shown to influence strongly the position of the CEP \cite{Fukushima:2008wg,Costa:2016vbb,Ferreira:2018sun}.

Experimentally, to gather evidences of the CEP’s location is one of the major goals in the HIC program, with relevant developments in the last years \cite{Aggarwal:2010cw,Abelev:2009bw,Adamczyk:2013dal,Aduszkiewicz:2015jna,Vovchenko:2017ayq,Adamczyk:2017iwn}. Presently, it is expected that the second Beam Energy Scan (BES-II) program at the Relativistic Heavy Ion Collider (RHIC) at Brookhaven National Laboratory, operating at intermediate collision energies (corresponding up to $\mu_B=400$ MeV) can lead to its discovery or, at least, constrain its location on the phase diagram. The future Compressed Baryonic Matter (CBM) experiment at the Facility for Antiproton and Ion Research (FAIR), in Darmstadt, Germany, the Nuclotron-Based Ion Collider Facility (NICA) in Dubna, Russia, and the Japan Proton Accelerator Research Complex (J-PARC) in Tokai, Japan, will extend this program to even lower collision energies, meaning even higher $\mu_B$ and lower temperatures (see \cite{Busza:2018rrf} for details).

Following along the lines of of previous works where isentropic trajectories were studied (see \cite{Ferreira:2017wtx,Ferreira:2018sun,Costa:2016wkj,Costa:2019bua}) here we analyse the effect upon these of the inclusion of eight quark interactions. Furthermore, contrarily too previous works, where a heat kernel expansion was used to derive the meson spectra (see \cite{Hiller:2008nu,Osipov:2012kk,Osipov:2013fka,Moreira:2014qna}), here we use a standard approach of expanding the Lagrangian to second order in the fields.

In HIC, the evolution of the fireball is accepted to be a hydrodynamic expansion of an ideal fluid thus being an isentropic process. This means that it will follow trajectories of constant entropy per baryon, $s/\rho_B$ (the so-called isentropes), in the phase diagram. For AGS, SPS, and RHIC, the values of $s/\rho_B$ are 30, 45, and 300, respectively \cite{Bluhm:2007nu}. LQCD simulations for the isentropic (2+1)-flavor equation of state (EOS) at these values of $s/\rho_B$  were presented in \cite{DeTar:2010xm,Borsanyi:2012cr}.

%PC: Removi o texto a seguir já que não vamos à velocidade do som

%Other important quantities for HIC are: the speed of sound of the system that is responsible for the collective acceleration of the fireball \cite{Rafelski:2015cxa}; and fluctuations of conserved charges that are encoded in quark-number susceptibilities. Together, they leave their imprint in final observables. 

The paper is organized as follows. In Section \ref{model_formalism} the extended PNJL model with eight quark interactions is formally introduced. In Section \ref{results_discussion} we present different parameter sets including eight quark interactions which are used to build different scenarios for the phase diagram. The isentropic trajectories are also presented and analysed with and without an extra term in the grand canonical potential which accounts for high momentum modes. Finally, in Section \ref{conclusions} conclusions are drawn.

%%%%%%%%%%%%%%%%%%%%%%%%%%%%%%%%%%%%%%%%%%%%%%%%%%%%%%%%%%%%%%%%%%%%%%%%%%%%%%%%
%%%%%%%%%%%%%%%%%%%%%%%%%%%%%%%%%%%%%%%%%%%%%%%%%%%%%%%%%%%%%%%%%%%%%%%%%%%%%%%%
\section{Model and formalism}
\label{model_formalism}

The Lagrangian density of the SU$(3)_f$ PNJL model including four, six and eight quark interactions can be written as \cite{Osipov:2005tq,Osipov:2006ns}:
\begin{align*}
\mathcal{L} & = 
\bar{q} \qty(i\slashed{D}-\hat{m}) q 
\\
& + \frac{G}{2}
[ \qty(\bar{q} \lambda_a q)^2 + 
(\bar{q} i \gamma^5 \lambda_a q)^2 ]
\\
& + 8 \kappa \qty[  
\det\qty( \bar{q} P_R q ) + 
\det\qty( \bar{q} P_L q )  ]
\\
& + 16 g_1 
\qty[ 
\qty( \bar{q}_i P_R q_j ) 
\qty( \bar{q}_j P_L q_i ) 
]^2
\\
& + 16 g_2 
\qty[ 
\qty( \bar{q}_i P_R q_j ) 
\qty( \bar{q}_j P_L q_k ) 
\qty( \bar{q}_k P_R q_l ) 
\qty( \bar{q}_l P_L q_i ) 
] 
\\
& 
-
\mathcal{U}\qty( \Phi , \bar{\Phi} ; T ) .
\numberthis
\label{NJL_Lagrangian}
\end{align*}
Where $q$ is a $N_f$-component vector in flavour space and $\hat{m}=diag\qty(m_u, m_d, m_s )$ is the quark current mass matrix. In this model the quark field is minimally coupled to a background gluonic field in the time direction, $A^0=g \mathcal{A}_a^0 \frac{\lambda_a}{2}$ through the covariant derivative, $D^\mu = \partial^\mu - i \delta_0^\mu A^0 $, $A_0 = -iA_4$ and $\mathcal{U}\qty( \Phi , \bar{\Phi} ; T )$ is the effective glue potential parametrized by the Polyakov loop\footnote{$\mathcal{P}$ is the path ordering operator.}:
\begin{align}
\Phi = 
\frac{1}{N_c} \; \underset{c}{ \tr} 
\mathcal{P} \exp 
\qty[ i \int_0^\beta \dd{\tau} A_4 \qty( \tau, \vec{x} ) ] .
\end{align}

In the interaction terms,  $\lambda^a$ ($a=1,2...8$) are the Gell-Mann matrices of the SU(3) group and $\lambda^0=\sqrt{\nicefrac{2}{3}} \mathds{1}$. The implicit sum and the determinant are to be carried out over flavour space. The chiral projector operators are defined as $P_{ \nicefrac{R}{L} } = (1 \pm \gamma^5)/2$. Finite density effects can be considered by including, in the lagrangian density, the term $\bar{q} \gamma^0\hat{\mu} q$, with  $\hat{\mu}=diag\qty(\mu_u, \mu_d, \mu_s )$ the chemical potential matrix.

The four quark scalar and pseudo-scalar interaction, is responsible for the dynamical breaking of chiral symmetry for a high enough coupling constant $G$.

The 't Hooft determinant interaction is introduced to explicit break the U$_A(1)$ symmetry. In three flavours, this interaction corresponds to a six quark interaction  which is known to destabilize the vacuum of the model making the potential unbounded by below \cite{Osipov:2005sp}.

The eight quark interactions were introduced to stabilize the ground state of the model \cite{Osipov:2005tq,Osipov:2006ns}. They constitute the most general spin-zero and chirally symmetry preserving interactions that can be introduced in the model without derivative terms. The first interaction term, with the $g_1$ coupling constant exhibits OZI violating effects.

The PNJL model is capable of describing the statistical confinement-deconfinement transition, with the breaking of $Z(N_c)$ symmetry using the Polyakov loop as an approximate order parameter\footnote{For pure glue theory, the Polyakov loop is an exact order parameter. In the confined phase, the boundary conditions of QCD are respected by the $Z(N_c)$ 
symmetry while in the deconfined phase it is broken.} 
. In the confined phase $\Phi \rightarrow 0$ while in the deconfined phase, $\Phi \rightarrow 1$.  The Polyakov loop effective potential, $\mathcal{U}\qty( \Phi , \bar{\Phi} ; T )$, has to be symmetric under $Z\qty(N_c)$ symmetry and reproduce its breaking at high temperatures.

In this work we will use the following polynomial potential
%\cite{Ratti:2005jh,Roessner:2006xn,Fukushima:2008wg}:
%\begin{align}
%\frac{\mathcal{U} \qty(\Phi,\bar{\Phi};T )}{T^4} & = 
%-\frac{1}{2} a\qty( T ) \bar{\Phi} \Phi + b\qty( T ) 
%\ln 
%\qty[ 
%1-6\Phi\bar{\Phi}+4(\Phi^3+\bar{\Phi}^3)-3(\Phi\bar{\Phi})^2
%],
%\label{eq:Polyakov.loop.potential}
%\\
%a\qty( T ) & = a_0 + a_1 \qty( \frac{T_0}{T} ) + a_2 \qty( \frac{T_0}{T} )^2; \,\,\,\,
%b\qty( T ) =  b_3 \qty( \frac{T_0}{T} )^3 .
%\end{align}
%Here, $a_0 = 3.51$, $a_1 = -2.47$, $a_2 = 15.2$ and $b_3 = -1.75$ \cite{Roessner:2006xn} obtained with $T_0=270$ MeV to reproduce lattice QCD result.
%\cite{Ratti:2005jh}
:
\begin{align}
\frac{\mathcal{U} \qty(\Phi,\bar{\Phi};T )}{T^4} & = 
-\frac{b_2\qty( T )}{2}\bar{\Phi}\Phi-\frac{b_3}{6}(\Phi^3+\bar{\Phi}^3)+\frac{b_4}{4}(\bar{\Phi}\Phi)^2
,
\label{eq:Polyakov.loop.potential}
\\
b_2\qty( T ) &=  a_0+a_1\qty( \frac{T_0}{T} )+a_2\qty( \frac{T_0}{T} )^2 +a_3\qty( \frac{T_0}{T} )^3.
\nonumber
\end{align}
Here, $a_0 = 6.75$, $a_1 = -1.95$, $a_2 = 2.625$, $a_3 = -7.44$, $b_3 = 0.75$ and $b_4 = 7.5$ obtained with $T_0=270$ MeV to reproduce lattice QCD data in the pure gauge sector \cite{Ratti:2005jh}. We opted to use this simple form of the Polyakov potential so as to better isolate the effect of eight quark interactions. Furthermore it should be noted that in our implementation of the thermodynamical potential, with the additional term to account for the high momentum modes (see below for discussion) the problem of the incorrect asymptotic value for the Polyakov loop does not occur \cite{Moreira:2010bx}.

Using the Lagrangian density given in Eq. (\ref{NJL_Lagrangian}), one can calculate the generating functional of the theory and relate it to the grand canonical potential, in order to study thermodynamical properties of the model. The presence of more than two quark interactions at the Lagrangian level renders an exact integration of the quark fields, impossible. In this work we will use the so called mean field approximation where quark bilinear operators, $\hat{ {O} }$, are expanded around their mean values, plus a small perturbation as: $\hat{ {O} } = \expval*{ {O} } + \delta\hat{ {O} }$. Terms of order superior to $ \delta\hat{ {O} } $ are neglected, effectively linearizing the Lagrangian.

Using the Matsubara formalism \cite{Kapusta:Book}, the MF grand canonical potential, $\Omega$, for the SU(3) NJL model within the 3-momentum regularization scheme, at finite temperature and chemical potential, for the $N_c=3$ case, can be written as:
\begin{align*}
\Omega \qty(T, \mu) & = 
\Omega_0  + 
U + 
\mathcal{U} \qty(\Phi,\bar{\Phi};T ) +
C \qty(T,\mu)
\\
& - 
6
\sum_{i} 
\int_0^\Lambda \frac{ \dd[3]{p} }{(2\pi)^3}  
\qty[ 
E_i +
\mathcal{F}\qty( T,\mu_i ) + 
\mathcal{F}^* \qty( T,\mu_i )  
] .
\numberthis
\label{NJLpot}
\end{align*}
with the sum made over $i=\{u,d,s\}$. The potential $U$ and the thermal functions $\mathcal{F} \qty( T,\mu_i )$ and $\mathcal{F}^* \qty( T,\mu_i )$, defined as:
\begin{align}
U & =
G \sum_{i} \sigma_i^2
+ 4 \kappa \prod_{i} \sigma_i
+ 3 g_1 \Big( \sum_{i} \sigma_i^2 \Big)^2
+ 3 g_2 \sum_{i} \sigma_i^4 ,
\end{align}
\begin{align*}
\mathcal{F} \qty( T,\mu_i ) = 
T \ln 
\Big[ 
1 & + 
3 \bar{\Phi} \e^{- \frac{1}{T} \qty(E_i-\mu_i)  } +
3 \Phi \e^{-\frac{2}{T} \qty(E_i-\mu_i )} 
\\
& + 
\e^{- \frac{3}{T} \qty(E_i-\mu_i) } 
\Big] ,
\numberthis
\label{Fthermal}
\end{align*}
\begin{align*}
\mathcal{F}^* \qty( T,\mu_i ) = 
T \ln 
\Big[ 
1 & + 
3 \Phi \e^{- \frac{1}{T} \qty(E_i+\mu_i) } + 
3 \bar{\Phi} \e^{- \frac{2}{T} \qty(E_i+\mu_i) } 
\\
& + 
\e^{- \frac{3}{T} \qty(E_i+\mu_i) }
\Big]   .
\numberthis
\label{F*thermal}
\end{align*}
with  $E_i=\sqrt{p^2+M_i^2}$ and $\expval*{\bar{q}_i q_i} = \sigma_i$ the quark condensate. The constant $\Omega_0$ is the thermodynamical potential in the vacuum, $\Omega_0=\Omega(T=0,\mu=0)$, ensuring that the vacuum pressure is zero.

%\begin{widetext}
%\begin{align*}
%\Omega \qty(T, \mu) & = 
%\Omega_0  + 
%U + 
%\mathcal{U} \qty(\Phi,\bar{\Phi};T )
% - 
%2 N_c 
%\sum_{i} 
%\int_0^\Lambda \frac{ \dd[3]{p} }{(2\pi)^3}  
%\qty[ 
%E_i +
%\mathcal{F}\qty( T,\mu_i ) + 
%\mathcal{F}^* \qty( T,\mu_i )  
%]
%+ C \qty(T,\mu) .
%\numberthis
%\label{NJLpot}
%\end{align*}
%with the sum made over $i=\{u,d,s\}$. The potential $U$ and the thermal functions $\mathcal{F} \qty( T,\mu_i )$ and $\mathcal{F}^* \qty( T,\mu_i )$, defined as:
%\begin{align}
%U & =
%G \sum_{i} \sigma_i^2
%+ 4 \kappa \prod_{i} \sigma_i
%+ 3 g_1 \Big( \sum_{i} \sigma_i^2 \Big)^2
%+ 3 g_2 \sum_{i} \sigma_i^4 ,
%\\
%\mathcal{F} \qty( T,\mu_i ) & = 
%T \ln 
%\qty[ 
%1 + 
%\e^{-3\qty(E_i-\mu_i)/T} + 
%N_c \bar{\Phi} \e^{-\qty(E_i-\mu_i)/T} + 
%N_c \Phi \e^{-2 \qty(E_i-\mu_i )/T} 
%] ,
%\label{Fthermal}
%\\
%\mathcal{F}^* \qty( T,\mu_i ) & = 
%T \ln 
%\qty[ 
%1 + \e^{-3\qty(E_i+\mu_i)/T} + 
%N_c\Phi \e^{-\qty(E_i+\mu_i)/T} + 
%N_c\bar{\Phi} \e^{-2\qty(E_i+\mu_i)/T} 
%]   .
%\label{F*thermal}
%\end{align}
%with $\expval*{\bar{q}_i q_i} = \sigma_i$ the quark condensate. The constant $\Omega_0$ is the value of the thermodynamical potential in the vacuum and $E_i=\sqrt{p^2+M_i^2}$.  
%\end{widetext}

For $i\ne j\ne k\in \{u,d,s\}$, the $i-$quark effective mass, $M_i$, is  given by the gap equation:
\begin{align}
M_i & = m_i 
- 2 G \sigma_i
- 2\kappa \sigma_j \sigma_k 
- 4 g_1 \sigma_i \sum_{j} \sigma_j^2
- 4 g_2 \sigma_i^3 .
\end{align}

Minimizing the thermodynamic potential with respect to $\sigma_i$, $\Phi$, $\overline{\Phi}$,
\begin{align}
\pdv{ \Omega }{\sigma_i} = 
\pdv{ \Omega }{ \Phi} = 
\pdv{ \Omega }{ \bar{\Phi} } = 0 ,
\label{thermodynamic.consistency.PNJL}
\end{align}
we can determine the value of these quantities for a given temperature and chemical potential.

The temperature and chemical potential dependent term, $C \qty(T,\mu)$, in the grand canonical potential is defined as (for the $N_c=3$ case):
\begin{align*}
C \qty(T,\mu) = & 
 - 
6
\sum_{i} 
\int_\Lambda^\infty \frac{ \dd[3]{p} }{(2\pi)^3}   
T \ln \qty[ 1 + \e^{-(\abs{p}+\mu_i)/T} ]
\\
& - 
6
\sum_{i} 
\int_\Lambda^\infty \frac{ \dd[3]{p} }{(2\pi)^3}   
T \ln \qty[ 1 + \e^{-(\abs{p}-\mu_i)/T} ] 
.
\numberthis
\label{CTmu_def}
\end{align*}
Here, $\mu$ is the quark chemical potential.

This contribution represents an additional pressure of massless quarks coming from the thermodynamics of the high momentum modes, with $\abs{\vec{p}}>\Lambda$. These higher momentum modes are missing from the regularized PNJL grand canonical potential, where all integrations are limited to the cutoff, $\Lambda$. Adding such a contribution to the thermodynamics is essential to get the correct high-temperature behaviour of the thermodynamics in effective models \cite{Skokov:2010wb,Herbst:2010rf}. 

Indeed, by deriving the grand canonical potential of the model by integration of the gap equations, such contribution appears naturally \cite{Moreira:2010bx}. In such case, when integrating the gap equations over the squared mass from $0$ to $M_i^2$ (with $M_i$ the dynamical mass of the quark of flavor $i$) a subtraction of the thermal functions, $\mathcal{F}\qty( T,\mu_i )$ and $\mathcal{F}^* \qty( T,\mu_i )$, evaluated at zero mass appears. This procedure can be viewed as the model dependent determination of how much we are deviated from the massless case (as such both the thermal functions at $M_i$ and the zero mass subtraction should be regularized). In order to reproduce the correct thermodynamic behavior we should add the baseline of the pressure of a gas of massless non interacting fermions. As this baseline is non-model dependent it should not be regularized. In the particular case of the 3-momentum cutoff this results in the cancellation of the massless parts only up until the cutoff thus originating the additional term.

The thermodynamical quantities are determined via the thermodynamical potential (see \cite{Costa:2010zw}). The pressure is given by $P(T,\mu_i)=-\Omega(T,\mu_i)$ while the density of the 
$i-$quark, $\rho_i$, and the entropy density, $s$, are derived from the pressure using:
\begin{align}
\rho_i(T,\mu)\,= \left. \pdv{P(T,\mu_i)}{\mu_i}\right|_{T}\,,
\\
s(T,\mu)\,= \left. \pdv{P(T,\mu_i)}{T}\right|_{\mu} .
\end{align}

%%%%%%%%%%%%%%%%%%%%%%%%%%%%%%%%%%%%%%%%%%%%%%%%%%%%%%%%%%%%%%%%%%%%%%%%%%%%%%%%
%%%%%%%%%%%%%%%%%%%%%%%%%%%%%%%%%%%%%%%%%%%%%%%%%%%%%%%%%%%%%%%%%%%%%%%%%%%%%%%%
\section{Results and discussion}
\label{results_discussion}

The PNJL model in the vacuum has seven free parameters in the isotopic limit, the light current quark mass, $m_u=m_d=m_l$, the strange quark current mass, $m_s$, the scalar couplings, $G$, $K$, $g_1$, $g_2$ and the regularization cutoff, $\Lambda$. These parameters can be fixed by reproducing  vacuum observables such as the experimental values of meson masses and weak decays.

At finite temperature and chemical potential, the Polyakov loop potential parameter $T_0$ is responsible for controlling the temperature scale of the deconfinement transition. Its original value of $T_0=0.270~\mathrm{GeV}$, is chosen in order to reproduce the pure glue deconfinement transition \cite{Boyd:1996bx}. In fact, some works suggest that this parameter should be $N_f$ and chemical potential dependent \cite{Schaefer:2007pw}. In this work this parameter was fixed by requiring that the crossover deconfinement\footnote{Defined using the inflection point in the Polyakov loop.} temperature of the model is $T^\phi_c=0.172~\mathrm{GeV}$. This value was chosen in order to be bounded by the lattice QCD results from \cite{Aoki:2009sc}.

Previous works have performed the parametrization of this model using the so-called heat kernel expansion \cite{Osipov:2001nx,Osipov:2001bj,Osipov:2000rg}, as was mentioned earlier, alongside a Pauli-Villars regularization \cite{Pauli:1949zm} with two subtractions in the integrand \cite{Volkov:1984fr}. In this work, to perform the parametrization, we calculate the vacuum meson masses using  the more common quadratic expansion of the effective action using the 3-momentum regularization \cite{Klevansky:1992qe}. For more details, see the \ref{appendix_MF_meson_masses}.

The approach to fix the parameters was the following: for different values of the coupling $g_1$ ($g_1 = \{0,$ $800,$ $1600,$ $2400\}~\mathrm{GeV}^{-8}$), the remaining six free parameters were found by requiring the model to reproduce the masses of the pion ($M_{\pi^\pm}=0.140~\mathrm{GeV}$), the kaon ($M_{K^\pm}=0.494~\mathrm{GeV}$), the eta prime ($M_{\eta'}=0.958~\mathrm{GeV}$) and $a^\pm_0$ ($M_{a^\pm_0}=0.960~\mathrm{GeV}$) mesons, the leptonic decays of the pion ($f_{\pi^+}=0.0924~\mathrm{GeV}$) and kaon ($f_{K^+}=0.093~\mathrm{GeV}$). Note that this last constraint is slightly deviated from the empirical value, $f_{K^+} /f_{\pi^+} = 1.1928(26) $ \cite{Tanabashi:2018oca}. Hence, each value of $g_1$ corresponds a different parametrization and, therefore, to a different model. The resulting parameter sets are displayed in Table \ref{ParameterSetsI}.

The coupling constants of the model must obey certain inequalities between each order in order to stabilize the vacuum of the model. These stability conditions were studied in \cite{Osipov:2005tq} and are given by:
\begin{align}
g_1>0, \quad g_2 > -\frac{2}{3}g_1 , \quad G > \frac{\kappa^2}{8 g_1} .
\end{align} 
In the parameter sets displayed in Table \ref{ParameterSetsI}, only the $g_1=0~\mathrm{GeV}^{-8}$ does not fulfil such stability conditions. We analyse this case nonetheless because it is the one which is closer to not considering eight quark interactions in the model.

\begin{table*}
\caption{
Parameter sets: current masses of the light ($m_l=m_u=m_d$), and strange quarks ($m_s$), couplings for the NJL ($G$), 't Hooft determinant ($\kappa$) and OZI violating eight quark ($g_1$), and non OZI violating eight quark ($g_2$) interactions, 3-momentum cutoff used in the regularization ($\Lambda$) and $T_0$ parameter used in the polynomial Polyakov potential. These were obtained fitting the masses of the pion ($M_{\pi^\pm}=0.140~\mathrm{GeV}$), the kaon ($M_{K^\pm}=0.494~\mathrm{GeV}$), the eta prime ($M_{\eta'}=0.958~\mathrm{GeV}$) and $a^\pm_0$ ($M_{a^\pm_0}=0.960~\mathrm{GeV}$) mesons, the weak decays of the pion ($f_{\pi^+}=0.0924~\mathrm{GeV}$) and kaon ($f_{K^+}=0.093~\mathrm{GeV}$) and a crossover deconfinement temperature (defined using the inflection point in the Polyakov loop) of $T^\phi_c=0.172~\mathrm{GeV}$ when using the polynomial Polyakov potential (see Eq. \ref{eq:Polyakov.loop.potential}). The OZI violating coupling $g_1$ is fixed  at the listed values (and marked with $ ^\ast$).
} 
\label{ParameterSetsI}
\begin{footnotesize}
%\begin{tiny}
\begin{center}
\begin{tabular*}{\textwidth}{@{\extracolsep{\fill}}l|cc|cccc|c|c@{}}\hline
\multicolumn{1}{c}{} &
\multicolumn{1}{c}{$m_l$ [MeV]} & \multicolumn{1}{c}{$m_s$ [MeV]} & 
\multicolumn{1}{c}{$G~\left[\text{GeV}^{-2}\right]$} &
\multicolumn{1}{c}{$\kappa~\qty[\text{GeV}^{-5}]$}&
\multicolumn{1}{c}{$g_1~\qty[\text{GeV}^{-8}]$} &
\multicolumn{1}{c}{$g_2~\qty[\text{GeV}^{-8}]$}&
\multicolumn{1}{c}{$\Lambda~\qty[\text{GeV}]$}&
\multicolumn{1}{c}{$T_0~\qty[\text{GeV}]$} \\
\hline
a) &$6.00209$ & $136.669$ & $12.5417$ & $-168.493$ & $   0^\ast$ & $1165.44$ & $0.576331$& $0.18257$ \\
b) &$6.00209$ & $136.669$ & $11.5154$ & $-168.493$ & $ 800^\ast$ & $1165.44$ & $0.576331$& $0.18271$ \\
c) &$6.00209$ & $136.669$ & $10.4891$ & $-168.493$ & $1600^\ast$ & $1165.44$ & $0.576331$& $0.18280$ \\
d) &$6.00209$ & $136.669$ & $ 9.4628$ & $-168.493$ & $2400^\ast$ & $1165.44$ & $0.576331$& $0.18264$ \\
\hline
\end{tabular*}
\end{center}
\end{footnotesize} 
\end{table*}

\begin{table*}
\caption{
Vacuum observables for each parameter set defined in Table \ref{ParameterSetsI}: effective quark masses ($M_l$ and $M_s$), light and strange quark condensates ($\expval{\bar{q}_l q_l}^{1/3}$ and $\expval{\bar{q}_s q_s}^{1/3}$) and masses of the $\eta$ ($M_\eta$), $\kappa^{\pm}$ ($M_{\kappa^{\pm}}$), $\sigma$ ($M_\sigma$) and $f_0$ ($M_{f_0}$) mesons.
} 
\label{OutputSetsI}
\begin{footnotesize}
%\begin{tiny}
\begin{center}
\begin{tabular*}{\textwidth}{@{\extracolsep{\fill}}l|cc|cc|cccc@{}}\hline
\multicolumn{1}{c}{} &
\multicolumn{1}{c}{$M_l$ [MeV]} & \multicolumn{1}{c}{$M_s$ [MeV]} & 
\multicolumn{1}{c}{$ \expval{\bar{q}_l q_l}^{1/3}~\qty[\text{MeV}]$} &
\multicolumn{1}{c}{$ \expval{\bar{q}_s q_s}^{1/3}~\qty[\text{MeV}]$}&
\multicolumn{1}{c}{$ M_{\eta}~\qty[\text{GeV}]$} &
\multicolumn{1}{c}{$ M_{\kappa^{\pm}}~\qty[\text{GeV}]$}&
\multicolumn{1}{c}{$ M_{\sigma}~\qty[\text{GeV}]$}&
\multicolumn{1}{c}{$ M_{f_0}~\qty[\text{GeV}]$} \\
\hline
a) & $443.3$ & $619.1$ & $-240.8$ & $-251.3$ & $0.508$ & $1.136$ & $0.868$ & $1.294$  \\
b) & $443.3$ & $619.1$ & $-240.8$ & $-251.3$ & $0.508$ & $1.136$ & $0.836$ & $1.289$ \\
c) & $443.3$ & $619.1$ & $-240.8$ & $-251.3$ & $0.508$ & $1.136$ & $0.800$ & $1.284$ \\
d) & $443.3$ & $619.1$ & $-240.8$ & $-251.3$ & $0.508$ & $1.136$ & $0.758$ & $1.281$ \\
\hline
\end{tabular*}
\end{center}
\end{footnotesize} 
\end{table*}

The reason to fix the $g_1$ coupling \textit{a priori}, is connected to the aforementioned works that used the heat kernel expansion to calculate the meson masses. In these works it was observed that increasing the value of the $g_1$ coupling had the effect of decreasing the predicted value for the $\sigma$ and $f_0$ mesons (the latter only slightly) while keeping the rest of the low-lying scalar and pseudoscalar meson spectra unchanged. The same conclusion was observed in our approach, see Table \ref{OutputSetsI}. We point out that the identification of the scalars with physical states is debatable (apart from the trivial quantum numbers matching). For instance, it could be argued that, due to the dubious identification of the $\sigma$ meson with a simple $\overline{q}q$ state, the lowest lying scalar states should in fact be identified with the $f_0(980)$ and the $f_0(1370)$ physical states. It should be noted, however, that the only scalar meson used in the fitting procedure was $M_{a^\pm_0} =(980 \pm 20)~\mathrm{MeV}$. The remaining scalars are outputs of the model.

Upon analysing Table \ref{ParameterSetsI} a very important feature of the inclusion of eight quark interactions in the model is evident: the rise of the $g_1$ coupling constant only affects the magnitude of four fermion interaction, $G$. The remaining vacuum parameters are not changed by increasing $g_1$ in agreement with the previous, already mentioned, heat kernel expansion based reports. Of course, increasing $g_1$ implies that we are changing the model, as such, a change in the deconfiment transition temperature at $\mu_B=0~\mathrm{GeV}$ for a fixed $T_0$ is expected, which means that the $T_0$ parameter should be different for each parameter set to ensure that $T^\phi_c=0.172~\mathrm{GeV}$ for every parameter set. The values for $T_0$ span, however, a surprisingly narrow range, see Table \ref{ParameterSetsI}, resulting in a effective independence of $T_0$ on the $g_1$ coupling choice.

\begin{figure*}[ht!]
%%%%%%%%%%%%%%%%%%%%%%%%%%%%%%%%%%%%%%
%%%%%%%%%%%%%%%%%%%%%%%%%%%%%%%%%%%%%%
\begin{subfigure}[b]{0.5\textwidth}
\includegraphics[width=\textwidth]
{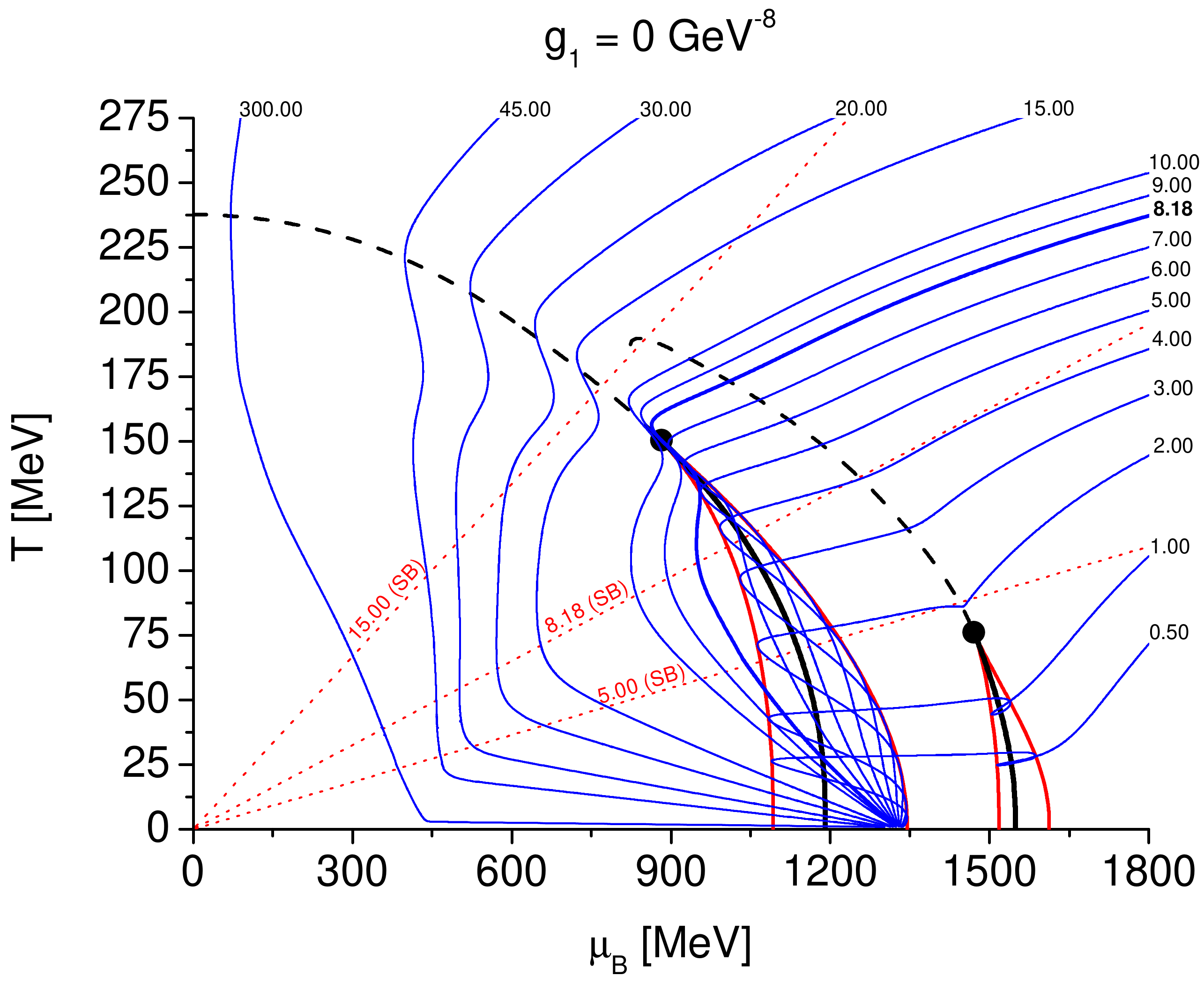}
\caption{}
\label{fig1a}
\end{subfigure}
\begin{subfigure}[b]{0.5\textwidth}
\includegraphics[width=\textwidth]
{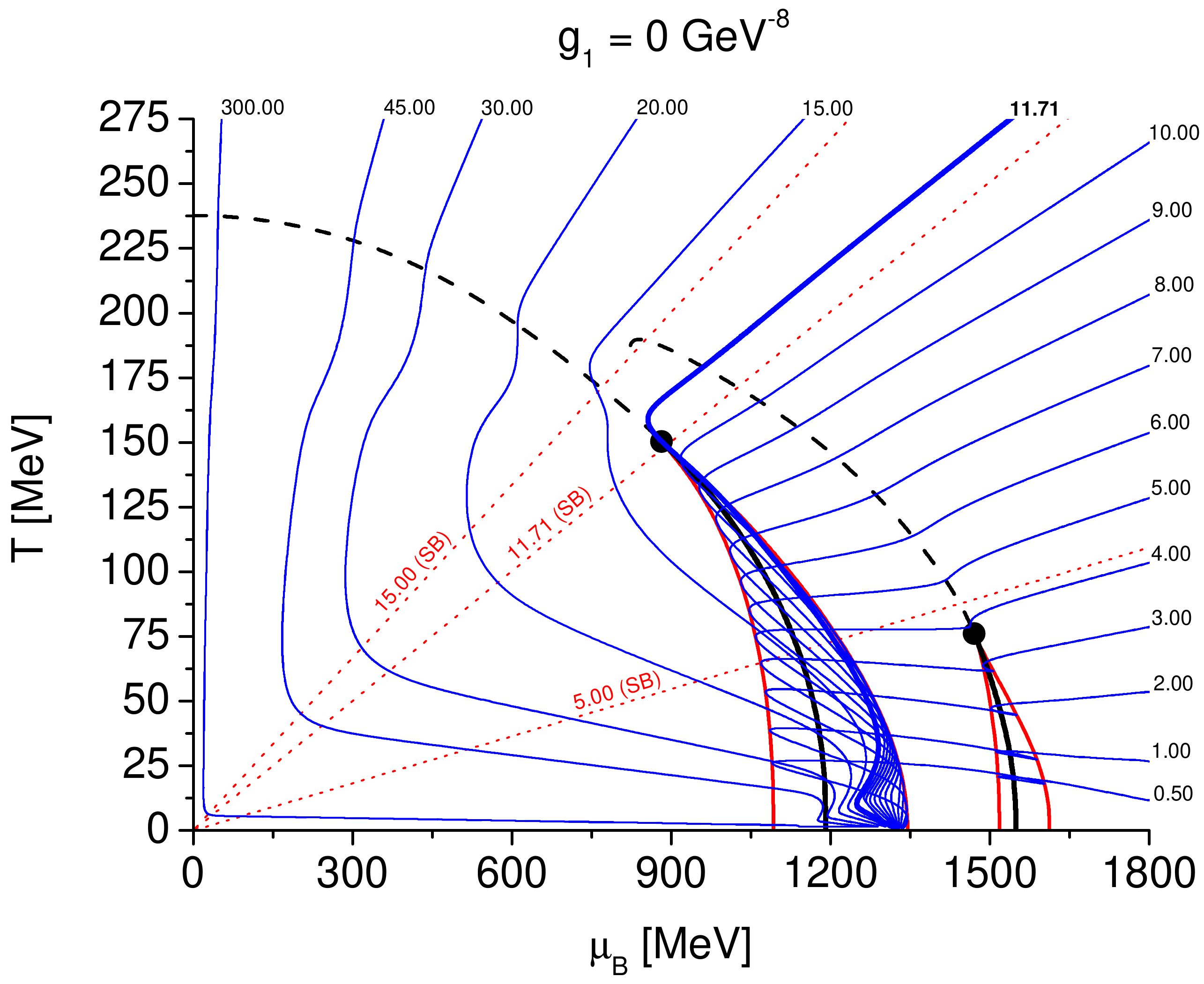}
\caption{}
\label{fig1b}
\end{subfigure}
%%%%%%%%%%%%%%%%%%%%%%%%%%%%%%%%%%%%%%
%%%%%%%%%%%%%%%%%%%%%%%%%%%%%%%%%%%%%%
%%%%%%%%%%%%%%%%%%%%%%%%%%%%%%%%%%%%%%
%%%%%%%%%%%%%%%%%%%%%%%%%%%%%%%%%%%%%%
\caption{The impact of the inclusion of $C\qty(T,\mu)$ in the thermodynamical potential upon the determination of the isentropic lines (constant entropy per baryon number) is illustrated above in panels \ref{fig1a} and \ref{fig1b} (without/with $C\qty(T,\mu)$  respectively). The chosen parametrization for this case study is parameter set a) from Table \ref{ParameterSetsI}. Full black lines correspond to the first-order transitions (ending in the CEPs marked with black circles) whereas dashed black lines correspond to crossover lines. The critical temperatures for the crossover lines are determined by using the inflexion point in $\sigma_i$ ($i=l$ for the leftmost line and $i=s$ for the rightmost) at fixed chemical potential. The spinodals are marked by full red lines. The isentropic lines are displayed in full blue lines (the chosen value of entropy per baryon number is displayed in the end of the curve). The thicker line corresponds to the isentropic line that goes through the CEP. Red dashed straight lines irradiating from the origin correspond to the Steffan-Boltzmann limit of the $s/ \rho_B=5,\,{\mathrm{CEP}}$, and 15 cases.}
\label{phase_diagram_g1_CTmu}
\end{figure*}

Lets now study the impact, in the phase diagram and isentropic trajectories, of the inclusion of $C\qty(T,\mu)$ in the thermodynamical potential of the model, using the a) parameter set in Table \ref{ParameterSetsI}, with $g_1 = 0~\mathrm{GeV}^{-8}$. Considering this parametrization, in Figure \ref{phase_diagram_g1_CTmu}, we present the first-order phase transition line, the spinodal lines, the CEP and several isentropic lines for symmetric matter i.e., $\mu=\mu_u=\mu_d=\mu_s=\mu_B/3$. In the left panel, Figure \ref{fig1a}, the model does not include the $C \qty(T,\mu)$ term while, in the right panel, Figure \ref{fig1b}, it is included in the thermodynamical potential.

The chiral phase first-order phase transition line is calculated using the Maxwell construction, using the Gibbs conditions of thermal, chemical and mechanical equilibrium. The critical temperatures for the crossover lines are determined using the inflexion point in $\sigma_i$ ($i=l$ for the light quarks and $i=s$ for the strange quark) at fixed chemical potential.

\begin{figure}[ht!]
\includegraphics[width=0.5\textwidth]
{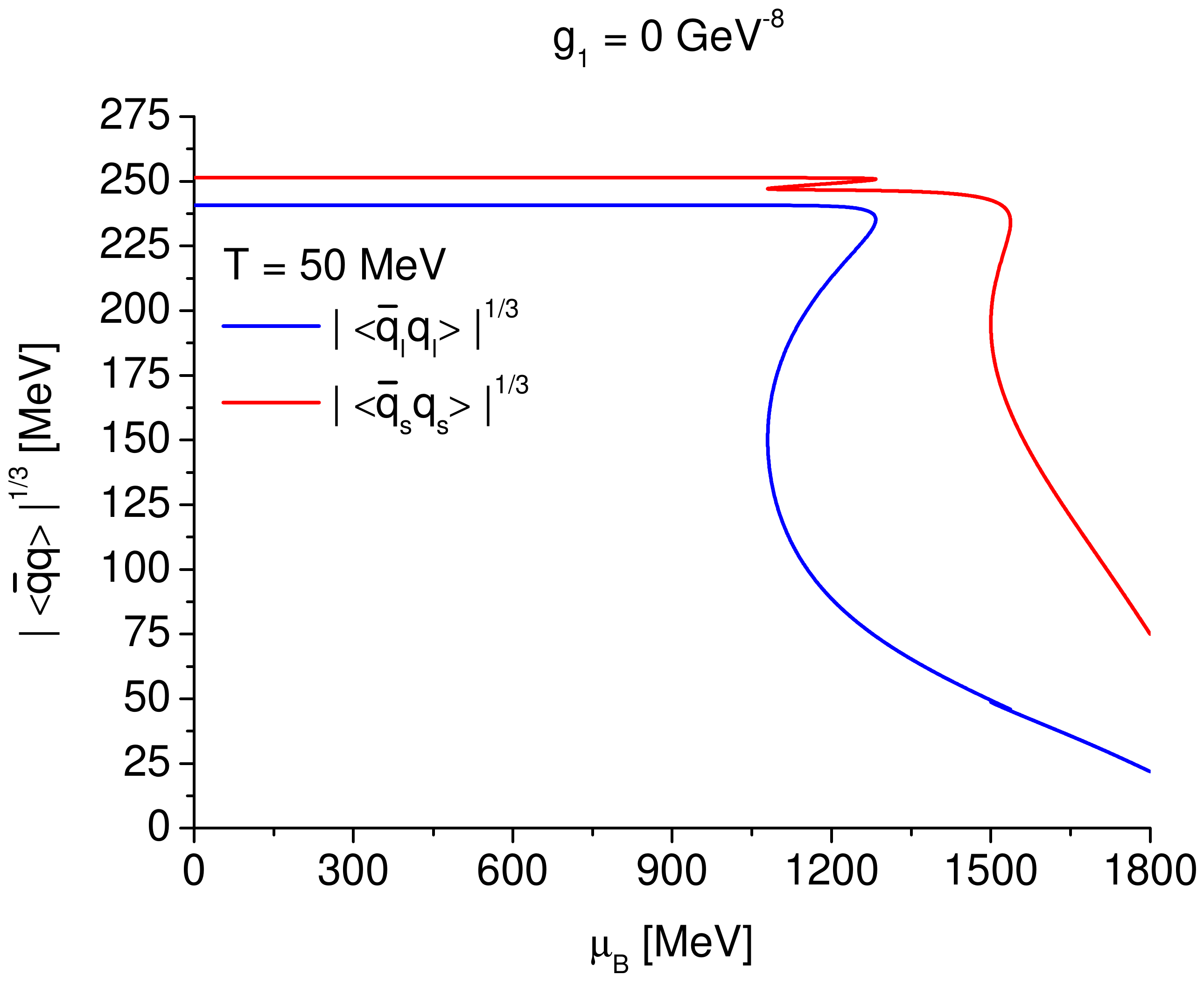}
\caption{Absolute value of the light and strange quark condensates ($\abs{\expval{\bar{q}_l q_l}}^{1/3}$ and $\abs{\expval{\bar{q}_s q_s}}^{1/3}$) as a function of the beryon chemical potential, $\mu_B$, for $T=50~\mathrm{MeV}$ considering $g_1=0~\mathrm{GeV^{-8}}$.}
\label{condensates_T50_g1_0}
\end{figure}

The first striking feature upon analysing these figures is the existence of two different first-order lines and CEPs: the leftmost one is due to the restoration of chiral symmetry in the light quark sector while the rightmost one is related with the restoration of chiral symmetry in the strange quark sector. The identification of these transitions with the restoration of chiral symmetry in the light and strange quarks sectors can be confirmed by observing Figure \ref{condensates_T50_g1_0}. In this figure we plot the light quark condensate and the strange quark condensate as function of the baryoninc chemical potential, $\mu_B$ for a fixed temperature of $T=50~\mathrm{MeV}$. The existence of two separate first order phase transitions and two CEPs will be discussed in more detail later.

Focusing in the effect of the inclusion of the  the $C \qty(T,\mu)$ term, the chiral critical region, i.e., the first-order phase transition line, spinodal lines and CEPs are not affect by this extra term. As a matter of fact, both critical regions are identical. The reason for this behaviour is quite evident when considering that the extra $C \qty(T,\mu)$ term does not depend on neither the condensate nor the Polyakov loop. Also, using the Gibbs conditions to define the first-order transition line, the chirally broken phase and the restored phase must be in thermal, chemical and mechanical equilibrium. In the latter requirement the pressure, for a given temperature and chemical potential, must be equal in both phases. Since the mass-independent $C \qty(T,\mu)$ term contribution to the pressure, is the same in both phases, it does not change the phase transition point.

On the other hand, the effect of the $C \qty(T,\mu)$ term in the isentropic trajectories, is completely different in each scenario. The general behaviour for the isentropic lines inside the critical region (for both, the light and strange first-order regions) can be informally described as bouncing back and forth between spinodals \cite{Costa:2016vbb}. In the previous section we claimed that the motivation to include such a term was to correctly reproduce the thermodynamic observables at finite temperature and chemical potential, such as the isentropic lines, by including, in the model the lacking higher momentum modes. For comparison purposes we also included the trajectories of constant entropy density of the quark-gluon gas, for three different scenarios: when the isentropic line crosses the first-order phase transition ($s / \rho_B = 5$), the CEP ($s / \rho_B = 8.18$ without $C \qty(T,\mu)$ and $s / \rho_B = 11.71$ with $C \qty(T,\mu)$) and the crossover transition ($s / \rho_B = 15$). In the Steffan Boltzmann limit, the isentropic trajectories for a gas of massless quarks and gluons in the $N_c=3$ and $N_f=3$ case (with $N_c$ and $N_f$ the number of colours and flavours, respectively) are given by (see for instance \cite{Allton:2005gk}):
\begin{align}
\frac{s\qty(T,\mu)}{\rho_B\qty(T,\mu)} = 
\frac
{ \frac{ \qty( 7 N_c N_f + 4 N_c^2 - 4 ) \pi^4 }{ 5 N_c N_f } T^3 + 3 \pi^2 T \mu^2  }
{ \pi^2 \mu T^2 + \mu^3 } .
\end{align}

In the case without $C \qty(T,\mu)$, Figure \ref{fig1a}, the isentropic trajectories for high values of temperature and chemical potential, where chiral symmetry is already restored, are completely different from the one expected in a free gas of massless quarks and gluons. In the case with $C \qty(T,\mu)$, Figure \ref{fig1b}, the isentropic trajectories have a very similar behaviour to what is expected from the Steffan Boltzmann limit case. Indeed, the same isentropic lines are parallel at high energies. It should be noted that a deviation from the ideal massless free quark-gluon gas is always expected due to the inclusion of a finite quark current mass (particularly in the case of the strange quark).

Besides the asymptotic differences between the isentropic trajectories with and without $C \qty(T,\mu)$, there are other important differences between both calculations. The isentropic lines that cross the light quark CEP ($s/\rho_B=8.18$ in the first case and $s/\rho_B=11.71$ in the second case) enters the critical region from the top , in the case without $C \qty(T,\mu)$, the isentropic line gets out from the critical region while, in the case with $C \qty(T,\mu)$, the isentropic line that cross the CEP remains bounded by the spinodal region of the chiral phase transition. Another difference is related with the larger isentropic trajectories with $s/\rho_B=15-300$: in the case with $C \qty(T,\mu)$ these lines are more spread in the phase diagram  and maintain a certain distance from one another while, in the other case they are closer together only getting more separated at high temperatures. In particular, the $s/\rho_B=300$ line in the case with $C \qty(T,\mu)$ is very close to the zero chemical potential axis for finite temperature.

\begin{figure*}[ht]
%%%%%%%%%%%%%%%%%%%%%%%%%%%%%%%%%%%%%%
%%%%%%%%%%%%%%%%%%%%%%%%%%%%%%%%%%%%%%
\begin{subfigure}[b]{0.5\textwidth}
\includegraphics[width=\textwidth]{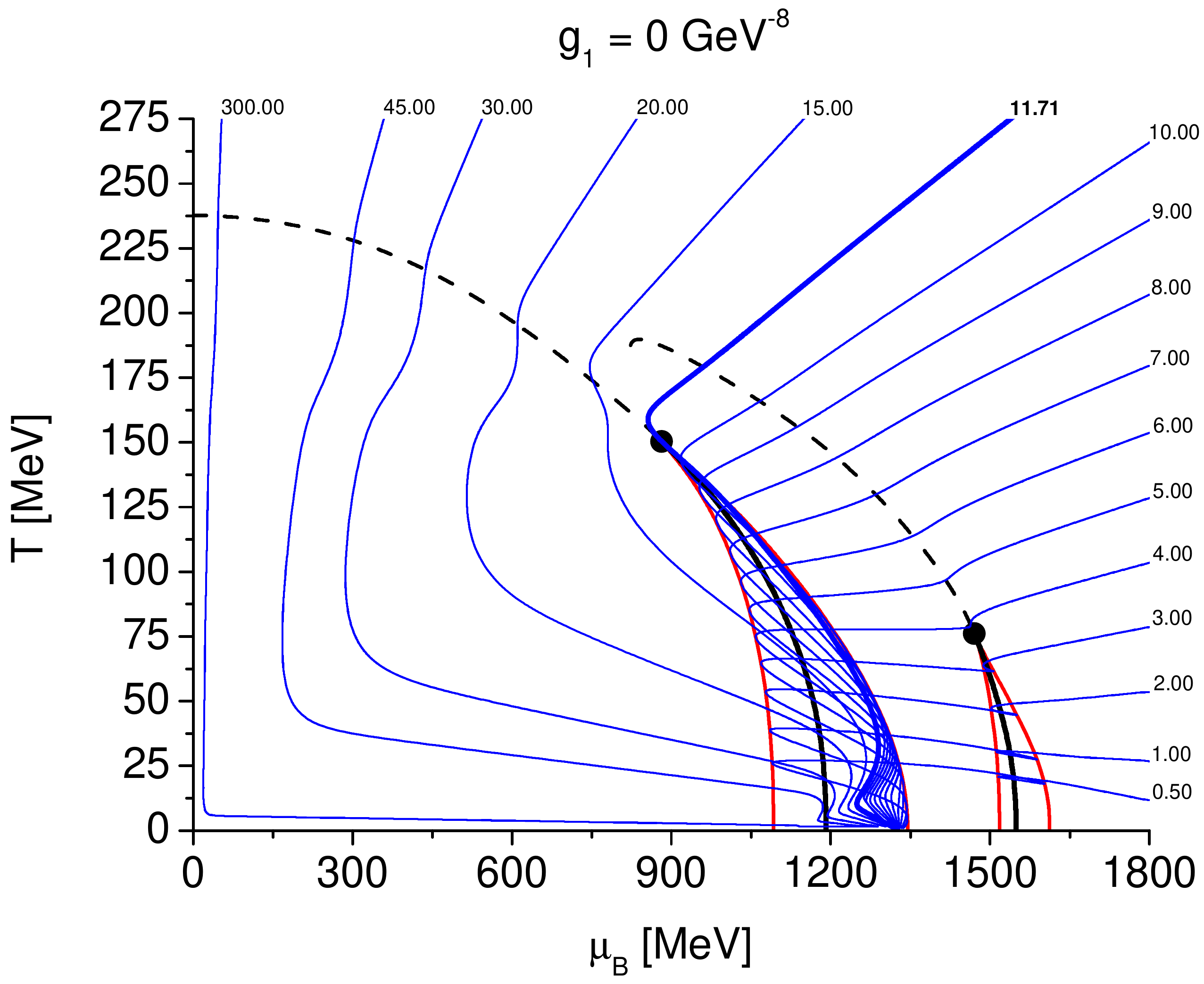}
\caption{}
\label{fig2a}
\end{subfigure}
\begin{subfigure}[b]{0.5\textwidth}
\includegraphics[width=\textwidth]{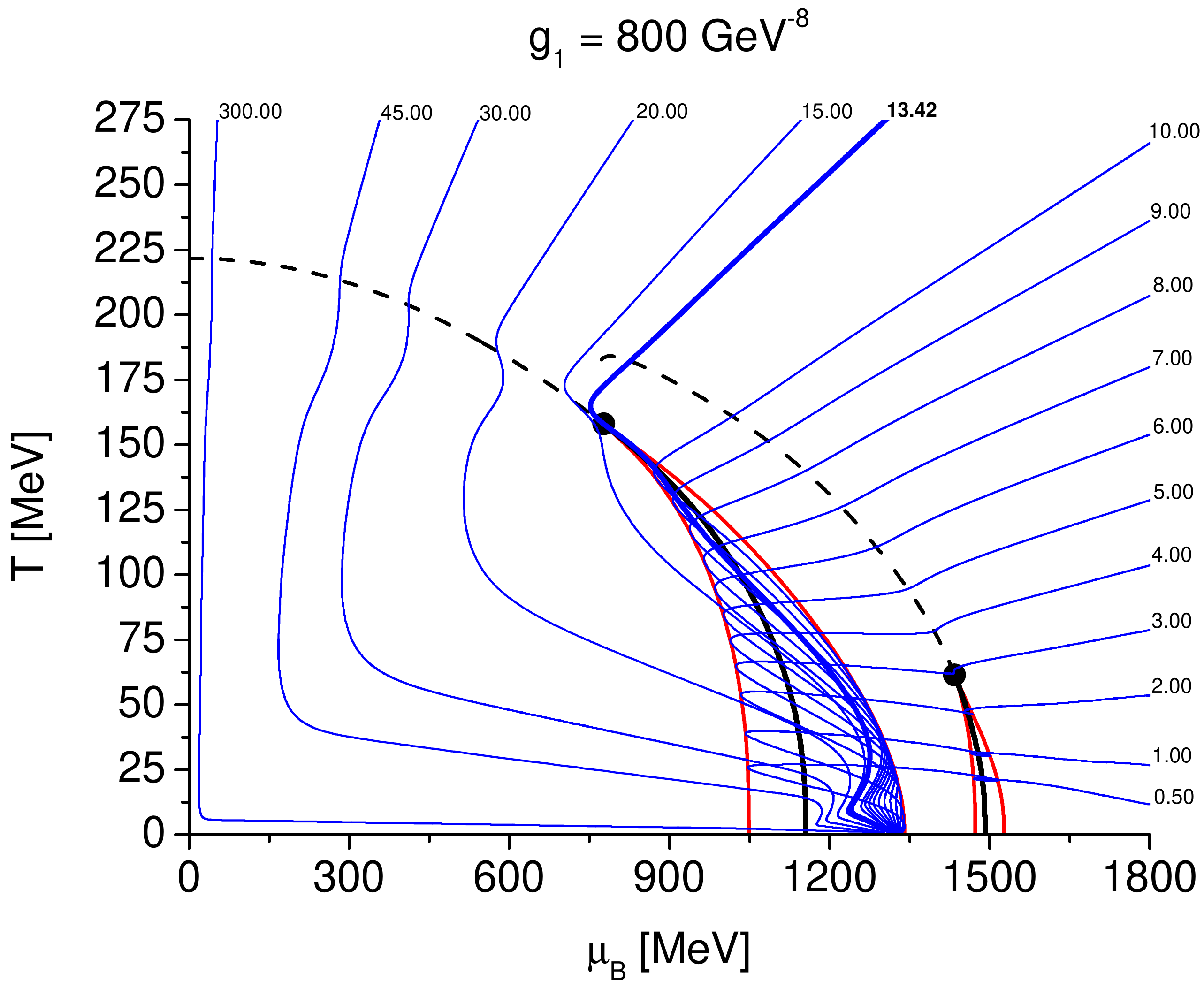}
\caption{}
\label{fig2b}
\end{subfigure}
%%%%%%%%%%%%%%%%%%%%%%%%%%%%%%%%%%%%%%
%%%%%%%%%%%%%%%%%%%%%%%%%%%%%%%%%%%%%%
\\ 
%%%%%%%%%%%%%%%%%%%%%%%%%%%%%%%%%%%%%%
%%%%%%%%%%%%%%%%%%%%%%%%%%%%%%%%%%%%%%
\begin{subfigure}[b]{0.5\textwidth}
\includegraphics[width=\textwidth]{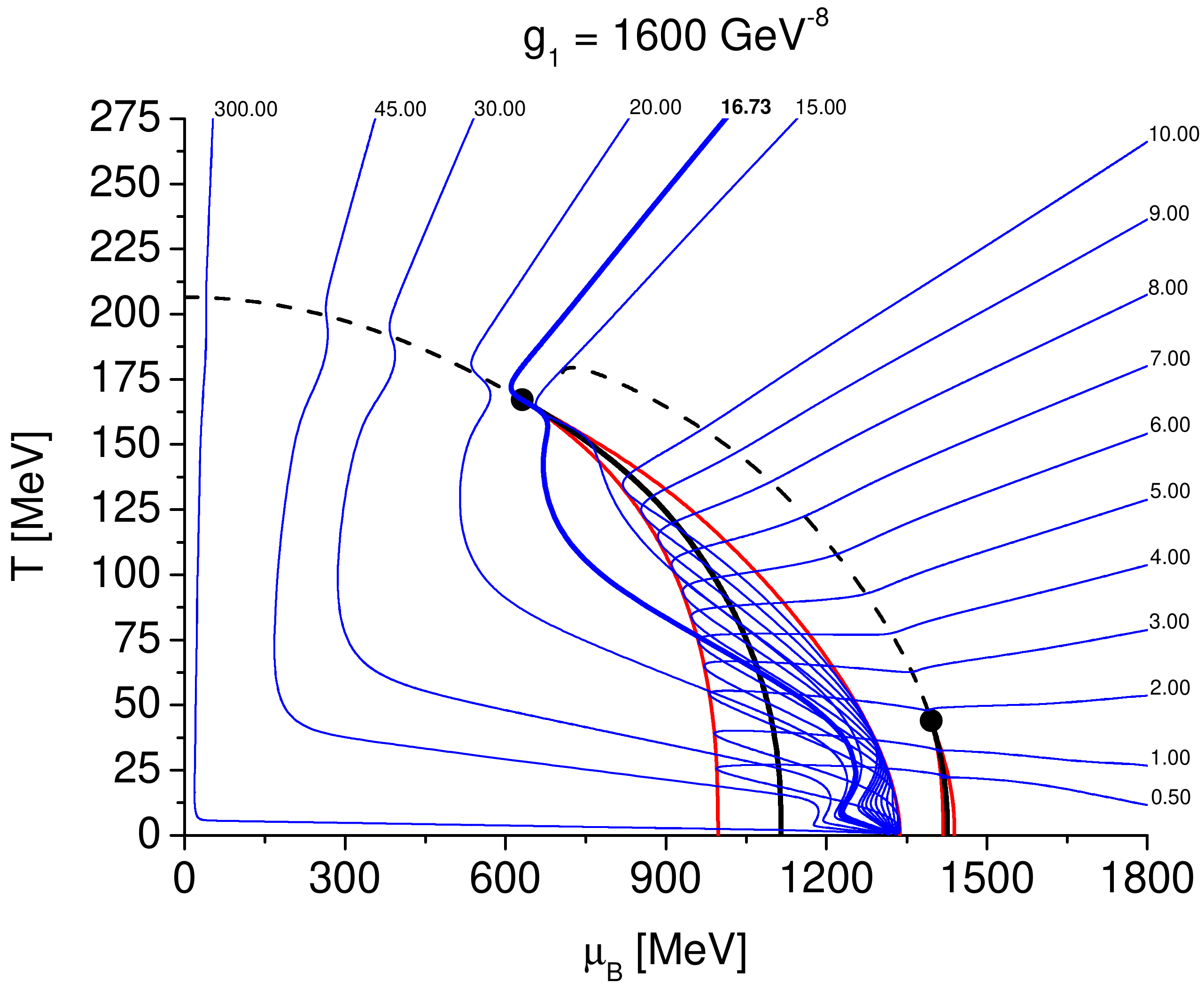}
\caption{}
\label{fig2c}
\end{subfigure}
\begin{subfigure}[b]{0.5\textwidth}
\includegraphics[width=\textwidth]{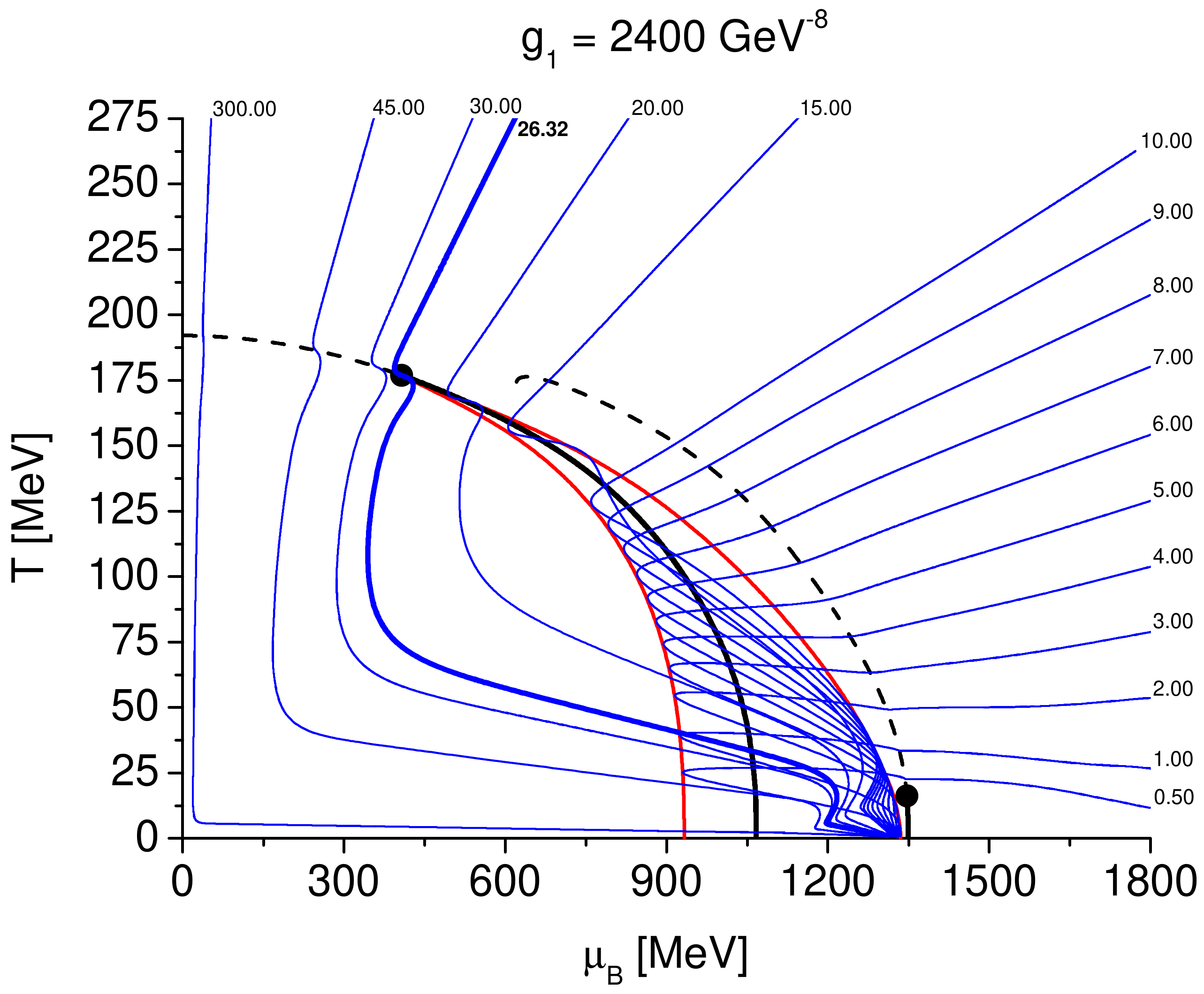}
\caption{}
\label{fig2d}
\end{subfigure}
%%%%%%%%%%%%%%%%%%%%%%%%%%%%%%%%%%%%%%
%%%%%%%%%%%%%%%%%%%%%%%%%%%%%%%%%%%%%%
\caption{The impact of the OZI violating eight quark interactions in the phase diagram and the isentropic lines determined by the model using parameter sets from Table \ref{ParameterSetsI} can be seen in the above panels. The line notation used is the same as in Figure \ref{phase_diagram_g1_CTmu}.}
\label{phase_diagram_g1}
\end{figure*}

We now turn our attention to the phase diagram and isentropic trajectories of models with different OZI violating coupling $g_1$, i.e., corresponding to each parametrization given in Table \ref{ParameterSetsI}. Such results can be observed in Figure \ref{phase_diagram_g1} where the $C \qty(T,\mu)$ term was included.

As already mentioned the most striking feature of these calculations is the presence of two CEPs, one related to the restoration of the light quarks and the other with the restoration of chiral symmetry of the strange quark. Multiple first-order phase transitions and CEP were observed when including the effect of a finite magnetic field \cite{Ferreira:2017wtx} In our calculation however, the strange CEP is present for all considered parameters sets.

The $\qty(\mu_B,T)$ coordinates of the light quark CEP for increasing values of $g_1=\qty{0,800,1600,2400}~\mathrm{GeV}$ are $\mathrm{CEP}_l=$ $\{\qty(882.1,150.4),$ $\qty(777.3,158.0),$ $\qty(631.3,167.1),$ $\qty(407.6,176.8)\}~\mathrm{MeV}$ while, for the strange CEP they are $\mathrm{CEP}_s=$ $\{\qty(1470.8,76.1),$ $\qty(1434.7,61.5),$ $\qty(1396.0,44.0),$ $\qty(1347.2,16.2)\}~\mathrm{MeV}$. The light quark CEP moves to larger temperatures and smaller chemical potentials. This was already observed in other calculations where eight quark interactions were incorporated \cite{Hiller:2008nu,Bhattacharyya:2010wp}. The behaviour of the strange CEP with increasing $g_1$, on the other hand, is very interesting: it moves to smaller baryon chemical and temperatures, contrary to the light CEP.

Focusing on the light quark phase transition, the first-order line and the leftmost spinodal line, at small temperatures, also moves towards smaller baryon chemical potentials. However the rightmost spinodal line almost does not move with increasing $g_1$. This means that the critical region, at smaller temperatures, gets larger with increasing $g_1$.  The crossover temperatures at $\mu_B =0$ also moves towards smaller temperatures with the increase of the OZI violating coupling, $T_c =\{ 237.7,$ $221.8 ,$ $206.5,$ $192.1 \}~\mathrm{MeV}$.

The hierarchy we found for the chiral and deconfinement transitions at $\mu_B =0$ is not consistent with LQCD calculations. From LQCD one expects the chiral transition temperature, $T_c^\chi\approx 156~\mathrm{MeV}$, to be lower temperature then the deconfinement temperature, $T_c^\phi~\approx 172~\mathrm{MeV}$. In this work we observed the opposite behaviour. The inability to correctly reproduce the hierarchy of the transitions of the PNJL model is a known limitation of the model we used. It should be noted that one can obtain the correct ordering but only at the cost of missing the scale at which they occur, see \cite{Hansen:2019lnf}. Increasing $T_0$ will drag both transitions to very high temperatures, although the order is correct. Also using the Entangled PNJL (a Polyakov loop dependence on the four quark coupling, see \cite{Sakai:2010rp}) one can make the two transitions occur simultaneously.

Regarding the strange crossover, it does not extend all the way to the $\mu_B=0$ axis, in fact, theres is a critical chemical potential for the appearance of this line. 

The absence of the strange crossover line for small chemical potentials is simply due to the disappearance of that particular inflection point. It  does not mean that there is no smooth transition, but only that the surviving inflection point is the one close to the crossover transition for light quarks (we choose not to add such line).

The $(\mu_B,T)$ position for the beginning of this line for the considered sets is, in increasing $g_1$ order: $\{\qty(823.5,187.5),$ $\qty(772.9,182.7),$ $\qty(707.1,177.7),$ $\qty(621.0,174.3) \}~\mathrm{MeV}$. Both the temperature and the chemical potential of the starting point of the strange crossover line are pushed towards lower values with increasing $g_1$ with the effect being more pronounced in the reduction of the latter.

When analysing the difference in the isentropic lines obtained with the chosen parametrizations one of the most interesting aspects is the small impact that the choice of $g_1$ has outside the critical region bounded by the spinodal lines. The delimitation of this region is however, as already mentioned, strongly influenced by the model choice.

For the critical region around the light quark chiral restoration transition we observe the following behaviour: starting from the point where it crosses the rightmost spinodal (coming in from the higher temperature/chemical potential region) it will continue more or less in the same path until it approaches the leftmost spinodal, there it turns around and continues until it approaches the rightmost spinodal, here it preforms another reversal and then, without reaching the leftmost spinodal, turns around one last time and follows the path dictated by the fact that all lines end at the critical chemical potential corresponding to the rightmost spinodal at vanishing temperature. 

It is worthwhile pointing out that not all lines follow this behaviour. Consider for instance the lines that go through the CEP: in the two cases with weaker OZI-violating eight quark interactions ($g_1=1600$ and $2400~\mathrm{GeV}^{-8}$, with $s/\rho_B=11.71$ and $s/\rho_B=13.42$, respectively) the lines stay inside the critical region, however, for the two cases with stronger OZI-violating eight quark interactions ($g_1=1600$ and $2400~\mathrm{GeV}^{-8}$), lines corresponding to an entropy per baryon number close (or equal) to $s_{CEP}$ ($s/\rho_B=16.73$ and $s/\rho_B=26.32$, respectively) in fact leave the critical region before reentering it at a lower temperature (ending as all others at the critical chemical potential of the rightmost spinodal at vanishing temperature). The smallness of the portion of the path spent inside the critical region by these $s_{CEP}$ isentropic lines for the strong $g_1$ cases results, in fact, in a strong similarity to the path of the lines that go through the crossover (with a $s$ slightly above that of the CEP). 

Also note that as the leftmost spinodals are shifted towards lower chemical potential (broadening the critical region) with increasing $g_1$, this means that lower chemical potentials can be reached inside this critical region.

For isentropic lines which also cross the critical region delimited by the spinodals resulting from the strange quark partial chiral restoration an additional loop appears with the line turning around at both spinodals. As this critical region becomes increasingly smaller with increasing $g_1$ this effect becomes almost imperceptible.

The presence of a first order phase transition in the diagram causes the absorption to the spinodal region of certain low temperature isentropic lines. For a fixed low baryonic chemical potential, the existence of a given isentropic line at high temperatures and its disappearance at low temperatures may be an experimental indication of the existence of a first order line in between the two high and low temperature regimes. Hence, the experimental observation of isentropic lines with smaller values of $s/ \rho_B$ then the ones that are currently being studied ($s/\rho_B=30,\,45,\,300$), could lead to an indication of a first order phase transition and indirectly the existence of the CEP. 

The isentropic lines that are currently under study ($s/\rho_B=30,\,45,\,300$), also display a bigger curvature when there is a CEP nearby. The observation of such a behaviour could also be an indirect hint of the existence of the CEP.

\section{Conclusions}
\label{conclusions}

We parametrized the vacuum PNJL model including four, six and eight quark interactions, in order to reproduce meson masses and leptonic decays using the 3-momentum regularization scheme. The meson masses were calculated using the usual quadratic expansion of the Lagrangian different from other approaches where the heat kernel expansion was used.

As observed in other works, \cite{Skokov:2010wb,Herbst:2010rf}, the inclusion of high momentum modes in the model is essential in order to calculate thermodynamic observables at high temperatures and chemical potentials. In this work such modes were added by including in the thermodynamic potential the temperature and chemical potential $C \qty(T,\mu)$ term, defined in Eq. (\ref{CTmu_def}).

The inclusion of eight quark interactions, as already reported in other works \cite{Hiller:2008nu,Bhattacharyya:2010wp}, pushes the CEP towards lower baryon chemical potentials and and higher temperatures. Such behaviour is controlled by the overall magnitude of the coupling related to the OZI violating interaction, $g_1$. Phenomenologically, different values of this coupling change only the masses of the $\sigma$ and, to a small extent, $f_0$ scalar mesons. Within our parametrizations the presence of a new first-order phase transition and CEP was obtained, related to the restoration of chiral symmetry of the strange quark. Contrary to the light quark CEP, the strange CEP moves to smaller temperatures with increasing $g_1$. The isentropic trajectories inside both the light and strange critical regions, have the same general behaviour (a bouncing back and forth between spinodals) while outside they are not affected very much by increasing values of this OZI violating coupling, $g_1$.

\begin{acknowledgements}
This work was supported by a research grant under project No. PTDC/FIS-NUC/29912/2017 (J.M.), funded by national funds through FCT (Fundação para a Ciência e a Tecnologia, I.P, Portugal)/ MCTES and co-financed by the European Regional Development Fund (ERDF) through the Portuguese Operational Program for Competitiveness and Internationalization, COMPETE 2020, by national funds from FCT under the IDPASC Ph.D. program (International Doctorate Network in Particle Physics, Astrophysics and Cosmology), with the Grant No. PD/\-BD/128234/\-2016 (R.C.P.), and under the Projects UID\slash FIS\slash 04564\slash 2019 and UID/FIS/04564/2020. The authors also acknowledge networking support by the COST Action CA15213 THOR (Theory of hot matter and relativistic heavy-ion collisions).

\end{acknowledgements}

%%%%%%%%%%%%%%%%%%%%%%%%%%%%%%%%%%%%%%%%%%%%%%%%%%%%%%%%%%%%%%%%%%%%%%%%%%%%%%%%
\appendix

\section{The Mean Field approximation and Meson Masses}
\label{appendix_MF_meson_masses}

We introduce the auxiliary scalar, $s_a$, and pseudoscalar field variables, $p_a$, written in terms of quark bilinear operators, $s_a=\bar{q} \lambda_a q$ and $p_a=\bar{q} i \gamma^5 \lambda_a q$, with indices $a=0,1,2...8$. Writing the Lagrangian density in terms of these new variables, yields:
\begin{align*}
\mathcal{L} & =
\bar{q} \qty(i\slashed{\partial}-\hat{m}) q 
\\
& 
+ \frac{G}{2} 
( s_a^2 + p_a^2 )
\\
& 
 + \frac{\kappa}{4} 
A_{abc} s_a ( s_b s_c - 3 p_b p_c )
\\
& + \frac{g_1}{4} 
( s_a^2 + p_a^2 )^2
\\
& + \frac{g_2}{8} 
\qty[ 
d_{abe} d_{cde} 
( s_a s_b s_c s_d + 2 s_a s_b p_c p_d + p_a p_b p_c p_d )
]
\\
& + \frac{g_2}{8} 
\qty(  
4 f_{abe} f_{cde} s_a p_b s_c p_d
) .
\numberthis 
\label{NJL_lag_auxfields}
\end{align*}
Here, $f_{abc}$ and $d_{abc}$ are the totally antisymmetric and  symmetric structure constants of the special unitary group SU(3), respectively. The constants $A_{abc}$ are defined as:
\begin{align*}
A_{abc} = 
\frac{2}{3} d_{abc} +
\sqrt{ \frac{2}{3} } 	
(
\delta_{a0} \delta_{b0} \delta_{c0} &-
\delta_{a0} \delta_{bc}
\\
& -
\delta_{b0} \delta_{ca} -
\delta_{c0} \delta_{ab}
) .
\numberthis
\end{align*} 

In order to derive the thermodynamical potential of the model we consider the mean field approximation. In this approximation, all quark interactions are transformed into quadratic interactions by introducing auxiliary fields whose quantum fluctuations are neglected and only the classical configuration contributes to the path integral i.e., the functional integration is dominated by the stationary point. A quark bilinear operator, $\operator$, can be written as its mean field value plus a small perturbation, $\operator = 
\expval*{ \operator } + \delta \operator$. To linearize the product of $N-$operators, terms superior to $(\delta \operator )^2$ must be neglected. Conveniently, the linear product between $N=n+1$ operators can be written using the following formula\footnote{\label{proof}This formula can be proved by induction.}:
\begin{align}
\prod_{i=1}^{n+1} \operator_i & =  
\qty[
\sum_{i=1}^{n+1} \frac{ \operator_i }{ \expval*{ \operator_i } }
- n  
]
\prod_{j=1}^{n+1} \expval*{ \operator_j } .
\label{linear_exp_Noperator}
\end{align}
The Lagrangian density can then be trivially linearized, the quadratic fermion term can be exactly integrated out and the grand canonical potential of the model can be derived to yield Eq. (\ref{NJLpot}).

The meson masses can be calculated by writing an effective Lagrangian, built by expanding the Lagrangian in Eq. (\ref{NJL_lag_auxfields}) up to second order in the auxiliary fields, \cite{Rehberg:1995kh}. Following the linear expansion of the Lagrangian, to build the quadratic expansion, terms superior to $(\delta \operator )^3$ must be neglected. More easily, the quadratic product between $N=n+2$ operators, with $n \geq 1$, can be written using the following formula\textsuperscript{\ref{proof}}
\begin{align*}
\prod_{i=1}^{n+2} \operator_i & =  
\bigg[
\frac{1}{2}
\sum_{i=1}^{n+2}
\sum_{j=1}^{n+2}
\frac{ \operator_i  }{ \expval*{ \operator_i } }
\frac{ \operator_j }{ \expval*{ \operator_j } }
\qty( 1 - \delta_{ij} )
\\
& \qquad - n  
\sum_{i=1}^{n+2}
\frac{ \operator_i }{ \expval*{ \operator_i } }
+\frac{n}{2} \qty(n+1)
\bigg]
\prod_{k=1}^{n+2} \expval*{ \operator_k } .
\numberthis
\label{quadratic_exp_Noperator}
\end{align*}

Having the quadratic expansion of the Lagrangian, the pseudoscalar and scalar inverse propagators are defined as the coefficient of the second order terms in the auxiliary fields. The pseudoscalar and scalar meson propagators are then given by:
\begin{align}
{G^{P}_{ab} (q)} 
& = 
\qty[ P_{ab}^{-1} - 4 \Pi^{P}_{ab}(q) ]^{-1}  ,
\\
{G^{S}_{ab} (q)} 
& = 
\qty[ S_{ab}^{-1} - 4 \Pi^{S}_{ab}(q) ]^{-1}  .
\end{align}
Here, the indices $a,b=0,1,2...8$. 

The pseudoscalar and scalar meson projectors, $P_{ab}$ and $S_{ab}$, with four, six and eight quark interactions, neglecting pseudoscalar condensates ($\expval{p_a}=0$), can be calculated to yield:
\begin{align*}
P_{ab}
& = 
G \delta_{ab} -
\frac{ 3\kappa }{2} A_{abc} \expval{s_c} +
g_1 \delta_{ab} \expval{s_c} \expval{s_c}
\\
& +\frac{g_2}{2} 
\qty( 
d_{abe} d_{cde} +
2 f_{dbe} f_{cae} 
)
\expval{s_c} \expval{s_d} ,
\numberthis
\\
S_{ab}
& = 
G \delta_{ab} +
\frac{3\kappa}{2} A_{abc} \expval{s_c} +
g_1
\qty(
\delta_{ab} \expval{s_c} \expval{s_c} +
2 \expval{s_a} \expval{s_b}
)
\\
& + \frac{g_2}{2}
\qty(
d_{abe} d_{cde} + 
d_{ace} d_{bde} + 
d_{ade} d_{cbe}
) 
\expval{s_c} \expval{s_d} .
\numberthis
\end{align*}

Using the diagonal matrices of SU$(3)_f$ and the identity, we can write the mean field values of the bilinear operators in the $0-3-8$ basis. One can switch to the quark flavour basis, $u-d-s$, doing a rotation as follows:
\begin{align}
\expval{s_a} = T_{ai} \sigma_i.
\end{align}
Here, the elements of the matrix $T_{ai}$ are given by:
\begin{align}
\qty(T_{ai})
=
\mqty[
\sqrt{2/3} & \sqrt{2/3} & \sqrt{2/3} 
\\ 
1 & -1 & 0
\\
1/\sqrt{3} & 1/\sqrt{3} & -2/\sqrt{3} 
] .
\end{align}

The polarization functions can be rotated between basis using,
\begin{align}
\Pi_{ab} = T_{ai} T_{bj} \Pi_{ij}.
\end{align}

The pseudoscalar and scalar polarization functions for two quarks with flavours $i$ and $j$, are given by \cite{Rehberg:1995kh}:
\begin{align*}
\Pi_{ij}^{P}(q) 
& = 
4
\qty[
I_i^{(1)} + I_j^{(1)} -
\qty( q^2 - \qty(M_i-M_j)^2 ) I_{ij}^{(2)}
]
,
\\
\Pi_{ij}^{S}(q) 
& = 
4
\qty[
I_i^{(1)} + I_j^{(1)} -
\qty( q^2 - \qty(M_i+M_j)^2 ) I_{ij}^{(2)}
] .
\end{align*}
Here,
\begin{align}
I_i^{(1)} & =
\frac{N_c}{4 \pi^2}
\int_0^\Lambda \dd{p}
\frac{p^2}{ E_i }
,
\\
I_{ij}^{(2)} & =
\frac{N_c}{4 \pi^2}
\CPV
\int_0^\Lambda \dd{p}
\frac{p^2}{ E_i E_j } \frac{ (E_i + E_j) }{ q^2 - (E_i+E_j)^2 }
.
\end{align}
Where $\CPV$ stands for the Cauchy principal value of the integral.

The mass of a given meson, $M_M$, and its decay width, $\Gamma_M$, can then be calculated by searching for the complex pole of its inverse propagator in the rest frame, i.e,
\begin{equation}
\qty[
G^{P/S}_{ab}
\qty(
M_M-i\frac{\Gamma_M}{2}, \vec{q} = \vec{0}
)
]^{-1} = 0.
\label{pole}
\end{equation}

The correspondence between the auxiliary pseudoscalar fields and the
physical pseudoscalar mesons can be performed using:
\begin{align}
\frac{\lambda_a \pi_a}{ \sqrt{2} } = 
\mqty[
\pi_u/\sqrt{2} & \pi^+ & K^+ 
\\ 
\pi^- & \pi_d/\sqrt{2} & K^0
\\
K^- & \bar{K}^0 & \pi_s/\sqrt{2}
] .
\end{align}
Where the pseudoscalar nonet was represented in the usual way. For the auxiliary scalar fields and the physical scalar fields, we use:
\begin{align}
\frac{\lambda_a \sigma_a}{ \sqrt{2} } = 
\mqty[
\sigma_u/\sqrt{2} & a_0^+ & \kappa^+ 
\\ 
a_0^- & \sigma_d/\sqrt{2} & \kappa^0
\\
\kappa^- & \bar{\kappa}^0 & \sigma_s/\sqrt{2}
] .
\end{align}
Using these correspondences the inverse propagator of a physical meson can be calculated using Eq. (\ref{pole}).

For the neutral mesons one must perform, as usual, a diagonalization of the quadratic contributions coming from the $0-3-8$ channels. In the isotopic limit, one therefore obtains the straightforward extension of the results from \cite{Rehberg:1995kh} to include the eight quark contributions.

\bibliographystyle{spphys}       % APS-like style for physics
\bibliography{draft_epja}   % name your BibTeX data base

%% Non-BibTeX users please use
%\begin{thebibliography}{}
%%
%% and use \bibitem to create references. Consult the Instructions
%% for authors for reference list style.
%\end{thebibliography}

\end{document}